\documentclass[aps,twocolumn,showpacs,groupedaddress,superscriptaddress,amsmath,amssymb]{revtex4}

\usepackage{graphicx}% Include figure files
\usepackage{dcolumn}% Align table columns on decimal point
\usepackage{bm}% bold math
\usepackage{epsfig,amsmath}

\def\vec#1{{\bf #1}}

\begin{document}

\title{Stability of superlubric sliding on graphite}

\author{Astrid S. de Wijn\footnote{e-mail:A.S.deWijn@science.ru.nl}}
\affiliation{Radboud University Nijmegen, Institute for Molecules and Materials, Heyendaalseweg 135, 6525AJ Nijmegen, the Netherlands}
\author{Claudio Fusco}
\affiliation{Radboud University Nijmegen, Institute for Molecules and Materials, Heyendaalseweg 135, 6525AJ Nijmegen, the Netherlands}
\author{Annalisa Fasolino\footnote{e-mail:A.Fasolino@science.ru.nl}}
\affiliation{Radboud University Nijmegen, Institute for Molecules and Materials, Heyendaalseweg 135, 6525AJ Nijmegen, the Netherlands}

\begin{abstract}
Recent AFM experiments have shown that the low-friction sliding of
incommensurate graphite flakes on graphite can be destroyed by
torque-induced rotations. Here we theoretically investigate the
stability of superlubric sliding against rotations of the flake. We
find that the occurrence of superlubric motion critically depends on
the physical parameters and on the experimental conditions: particular
scan lines, thermal fluctuations and high loading forces can destroy
the stability of superlubric orbits. We find that the optimal
conditions to achieve superlubric sliding are given by large flakes,
low temperature, and low loads, as well as scanning velocities higher than those
used in AFM experiments.


\end{abstract}

\maketitle

\section{Introduction}

Recent years have witnessed a surge of interest in understanding the
microscopic origin of friction as a result of the increased control in surface preparation,
the developments of local probes like the Atomic Force Microscopes (AFM)
and Scanning Tunneling Microscopes (STM)  and due to the interest for possible applications in nanotechnology.
One of the goals of this research is to understand whether extremely 
low friction can be obtained by an appropriate choice of the sliding conditions. 
This paper examines theoretically the sliding of graphite flakes on a graphite substrate, one of the prototype systems in this field.
For this system, it has recently been shown that the low friction 'superlubric' sliding reported previously for flakes with incommensurate contact with the
 substrate~\cite{Dienwiebel2004}
is always destroyed by rotations of the sliding flake~\cite{Dienwiebel2008}, leading to a locking in
a commensurate state with high friction and slip-stick behavior. Numerical simulations~\cite{Dienwiebel2008} carried out for the experimental conditions (extremely low velocities, about 30~nm/s) confirm this finding. It is intriguing to ascertain whether there might be conditions that avoid the rotation and locking in the high-friction commensurate orientation.

Some important concepts of friction at the atomic scale are based on the Frenkel Kontorova (FK) model~\cite{FK} that describes the sliding surface as a harmonic chain of lattice spacing $a$ in interaction with a rigid periodic substrate with period $b$. 
For incommensurate values of the  ratio $a/b$, Peyrard and Aubry~\cite{Aubry}
have shown that, below a critical value of the coupling to the periodic potential, the chain can be
displaced on the substrate by an infinitesimally small force, namely the system displays a vanishing
{\it static} friction force. 
Later, Shinjo and Hirano\cite{Shinjo}  predicted that for incommensurate contacts also the {\it kinetic} friction would vanish and called this effect {\it superlubricity}. 
The experimental  STM\cite{Hirano} and AFM studies~\cite{Dienwiebel2004}
showing a drop of the
friction force in going from commensurate
to incommensurate contacts seemed to confirm the prediction of superlubricity. 
Theoretical work\cite{Consoli} has shown that 
the prediction of frictionless sliding also at high velocities of Ref.~\cite{Hirano}
is oversimplified and does not apply in general, although dissipative mechanisms become less and less effective in the limit of vanishing velocities. Moreover the term superlubricity has been criticized in
several papers \cite{Weiss,Socoliuc} because it suggests a transition to zero friction which can
be compared to superfluidity or superconductivity, whereas there is no threshold
value of the velocity below which the kinetic friction vanishes. Nevertheless, the
term superlubricity has become very popular and is used to describe low friction in
the quasistatic limit accessible by AFM.

Here we study the driven dynamics of a finite graphite flake on a graphite surface.
The flake-surface interaction is modeled with a realistic static potential
but vibrations of the flake are not taken into account and those of the substrate are represented by an effective friction coefficient proportional to velocity.
This defines a deterministic non-linear dynamical system with four degrees of freedom that can be
studied by numerical simulations
and approximate analytical models, allowing us to study the stability of superlubric sliding.

For a commensurate contact, we always find a stick-slip behavior with
high friction.
Conversely, for an incommensurate contact we find two types of
qualitatively different behavior.
After an initial short period, the flake either rotates and locks into a
commensurate orientation or
it remains incommensurate and slides with extremely low friction.
This behavior is critically dependent on the initial conditions, as
expected for a strongly nonlinear
problem.
A simple dynamical system which captures the essential
physics and for which the stability analysis can be done analytically
explains the observed behavior.  We then examine
by numerical simulations
the stability of the periodic orbits corresponding to incommensurate
sliding against thermal
fluctuations and other perturbations.

In Sec.~\ref{sec:model} we describe the model of the structure and interactions and the details of the numerical simulations.
In Sec.~\ref{sec:per_orbits} we show that periodic orbits corresponding to either commensurate or incommensurate sliding appear for different initial conditions.  
In Sec.~\ref{sec:stability} we propose a simplified model for which we can perform analytically the stability analysis of these orbits.
The robustness of the stability of periodic orbits against different types of perturbations is presented in Section\ref{sec:robustness}. Finally we conclude with a summary and perspectives in Section\ref{sec:conclusions}. 

\section{\label{sec:model}Model}

\begin{figure}
\includegraphics[width=8.4cm]{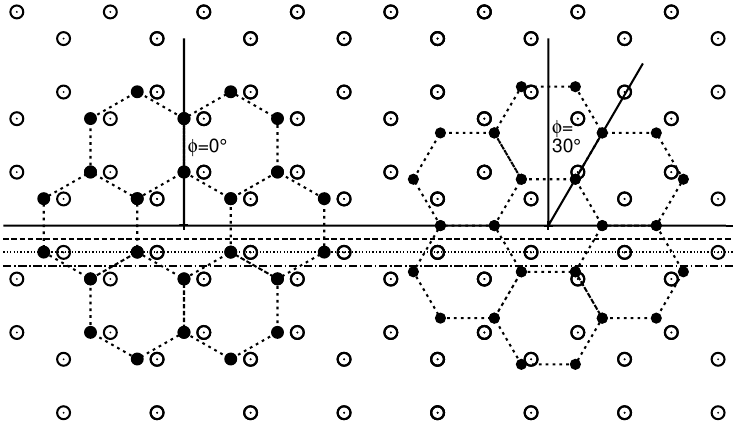}
\caption{
Top view of the geometry of a graphite flake of 24 atoms on the substrate, in a commensurate orientation (left, mismatch angle $\phi = 0$) and incommensurate orientation (right, $\phi = 30^\circ$).
The open circles represent substrate atoms, while the closed circles are flake atoms.
The scan lines used in this paper are along the $x$-axis and shown from top to bottom: scan line~1 (solid line), 2 (dashed line), 3 (dotted line), and~4 (dot-dashed line).
The scan lines are separated by a distance $a/4$.
Due to the symmetry of the lattice, the range between scan lines~1 and~4 fully describes all scan lines in this direction.  The scan line at a distance $a/4$ below scan line~4 is again equivalent to scan line~3.
Note that in a symmetric hexagonal flake, the center of mass does not correspond to the position of an atom.
}
\label{fig:geometry}
\end{figure}

We study the dynamics of rigid graphite flakes, lying in the $x-y$ plane parallel to the substrate as shown in Fig.~\ref{fig:geometry}.
Atoms are kept at the equilibrium inter-atomic spacing $a=1.42$~\AA\ in a hexagonal lattice for both the flake and substrate.
By changing the orientation of the flake onto the hexagonal
substrate the contact is either commensurate (Fig.~\ref{fig:geometry}, left) or incommensurate (Fig.~\ref{fig:geometry}, right).
We consider only rotations around the $z$ axis that keep the flake parallel to the substrate.
The center of mass of the flake is pulled along the indicated scan lines by a support moving at constant velocity $\vec{v}_\mathrm{s}=(v_\mathrm{s},0,0)$. 
The flake therefore has 4 degrees of freedom: the coordinates of the center of mass, $\vec{r} = (x, y, z)$ and the orientation $\phi$.
The corresponding velocities are $\vec{v} = (\vec{v}_x, \vec{v}_y, \vec{v}_z)$ and $\omega$.  The phase space has 8 dimensions.

We calculate the force and the torque acting on the center of mass from the interaction that each atom in the flake has with each atom in the substrate. 
The total potential energy of the flake due to interactions with atoms of the substrate can be written as
\begin{align}
V(\vec{r}, \phi) = \sum_{i} \sum_{j} V_\mathrm{C}(|\vec{r}_i-\vec{R}_j|)~,
\end{align}
where $i$ goes over all flake atoms, and $j$ over all substrate atoms and $V_\mathrm{C}(r)$ is the interaction between one flake atom and one substrate atom at distance $r$.
The positions of the substrate atoms $\vec{R}_j=(X_j,Y_j,Z_j)$ are given by a hexagonal lattice,
and the positions of flake atoms $\vec{r}_i=(x_i,y_i,z_i)$ are functions of the position of the center of mass $\vec{r}=(x,y,z)$ and of the orientation angle $\phi$ (see Fig.~\ref{fig:geometry}).
In the simulations described in this paper, we use the atom-atom interaction potential $V^\mathrm{LR}(r)$ of Ref.~\cite{Los2003} that describes non-bonded interactions of carbon.
The potential has a range of 6~\AA.

The support representing the AFM cantilever drives the flake, with a force given by
\begin{align}
F_{\mathrm{s}} (\vec{r},t) = - c \left(\begin{array}{l}x-x_\mathrm{s}(t)\\y-y_\mathrm{s}(t)\\0\end{array} \right) + \left(\begin{array}{l}0\\0\\-F_\mathrm{load} \end{array}\right)~,
\end{align}
where $t$ is the time, $(x_\mathrm{s},y_\mathrm{s},z_\mathrm{s}) = (x_\mathrm{s}(0) + v_\mathrm{s} t,y_\mathrm{s}(0),z_\mathrm{s})$ is the position of the support, $c$ (=~1~nN/nm) is the coupling constant between the support and the center of mass of the flake,
and $F_\mathrm{load}$ is the load force in the negative $z$ direction.
The coupling to the phonon modes of the substrate can be modeled by a viscous friction term that dampens the motion of the flake, with a force and torque given by
\begin{align}
F_{\mathrm{f}}(\vec{v}) = - \gamma M \vec{v}~,\\
T_{\mathrm{f}}(\omega) = - \gamma I \omega~,
\end{align}
where $M$ is the total mass of the flake, $I$ is the moment of inertia for rotations around the center of mass along the $z$-axis, and $\gamma$ (=~1/ps) is the viscous friction constant.
Note that for a rigid flake, the damping of the linear velocity directly determines the damping of both the centre of mass and the rotation.

The equations of motion are:
\begin{align}
\label{eq:motion1}
M \ddot{\vec{r}} & = -\frac{\partial V(\vec{r},\phi)}{\partial \vec{r}} + F_{\mathrm{s}} (\vec{r},t)+F_{\mathrm{f}}(\vec{v}) ~,\\
I \ddot\phi &= - \frac{\partial V(\vec{r},\phi)}{\partial \phi} + T_{\mathrm{f}}(\omega)~.
\label{eq:motion2}
\end{align}

The rotational symmetry of the flake implies that
\begin{align}
\label{eq:symmetries1}
V(\vec{r},\phi) = V(\vec{r},\frac\pi3+\phi)~.
\end{align}
and the periodicity of the substrate gives
\begin{align}
\label{eq:symmetries2}
V(\vec{r},\phi) = V(\vec{r}+\vec{a},\phi)~,
\end{align}
where $\vec{a}$ is any vector which generates a translation under which the lattice is invariant.
The flake-substrate system also has symmetry for reflections in the $yz$-plane
\begin{align}
\label{eq:symmetries3}
V(\vec{r},\phi) = V((-x,y,z), \pi-\phi)~.
\end{align}
In our numerical simulations, we solve the equations of motion using the velocity-Verlet algorithm with damping
and whenever the temperature is nonzero, a Langevin noise term is added~\cite{Dienwiebel2008,vverletspullie}.

\section{\label{sec:per_orbits}Periodic Orbits}

\begin{figure}
\includegraphics[width=8.4cm]{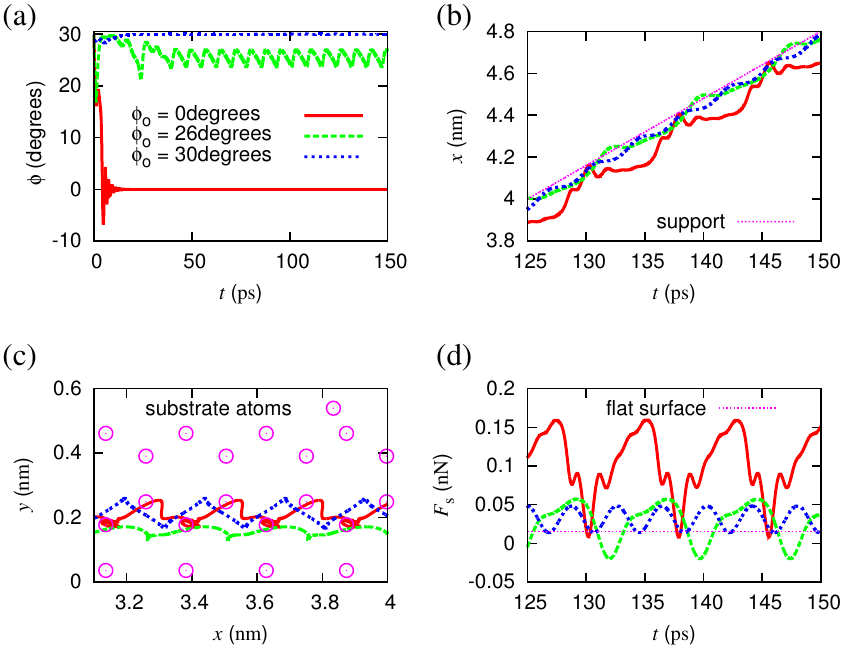}
\caption{
(Color online)
Three typical trajectories for a 24-atom flake subjected to $F_\mathrm{load} = 20~\mathrm{nN}, v_\mathrm{s} = 32~\mathrm{m/s}$.
All three converge to stable periodic orbits at approximately constant $\phi \approx \phi_0$.
The trajectories converging to $\phi_0\approx0^\circ$ and $\phi_0\approx26^\circ$ are for scan line 3, but have different initial angular velocity, and the trajectory converging to $\phi_0\approx30^\circ$ is for scan line 4.
From left to right and top to bottom: (a) the mismatch angle as a function of time, (b) the position as a function of time once the trajectories have converged to the periodic orbits, (c) the trajectories on the surface for the same interval, and (d) the friction force.
}
\label{fig:sometrajectories}
\end{figure}

The solutions of Eqs.~(\ref{eq:motion1}) and~(\ref{eq:motion2}) at $T=0$ are strongly dependent on the initial conditions, due to the nonlinearities of the interaction forces.
In Fig.~\ref{fig:sometrajectories} (top left) we show two trajectories obtained for exactly the same conditions (same load, support velocity, and scan line) apart from different initial angular velocity.
We can see that starting from an orientation  near the incommensurate orientation, $\phi$ either drops to the commensurate $\phi=0$ value, or oscillates around approximately $26^\circ$.
A similar trajectory on another scan line converges to $30^\circ$.
The orientation converges to a stable value within a few lattice periods.
This result shows that several periodic orbits may be stable.
The corresponding behavior of $x(t)$, shown in Fig.~\ref{fig:sometrajectories} (top right), for the commensurate case $\phi=0$ is step-like, which is typical of stick-slip motion.
For the incommensurate cases $\phi=26^\circ,30^\circ$, the flake follows the support closely.
The difference between commensurate and incommensurate orbits is also evident by looking at the trajectory in the $xy$-plane, shown at the bottom left of Fig.~\ref{fig:sometrajectories}.
In the case of $\phi=0$ the centre of mass jumps quickly from one lattice site to another, where it performs some oscillations before jumping again.
The incommensurate motion at the same scan line is smoother and the orbit at $\phi=30^\circ$ performs a regular zig-zag motion.
The lateral force, also displayed in Fig.~\ref{fig:sometrajectories} (right bottom), which shows stick-slip motion for the commensurate trajectory, drops for $\phi=26^\circ$ and $\phi=30^\circ$ to an average friction force close to that of a flat surface ($\gamma M v_\mathrm{s} = 0.0153~\mathrm{nN}$), $0.0278~\mathrm{nN}$ and $0.0316~\mathrm{nN}$ respectively.
The friction of the commensurate flake, by comparison, is large, $0.1018~\mathrm{nN}$.

\begin{figure}
\includegraphics[width=8.4cm]{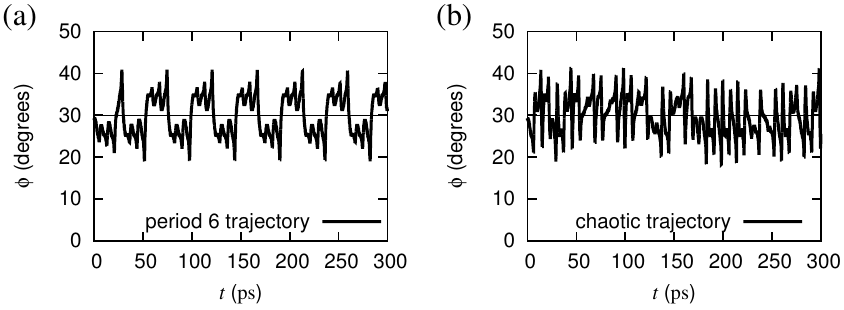}
\caption{
Examples of more complicated trajectories of a 24-atom flake: (a) a periodic trajectory with a longer period, in this case 6 lattice periods, for $F_\mathrm{load} = 30 \mathrm{nN}$ at scan line 3 and (b) a chaotic trajectory for for $F_\mathrm{load} = 40 \mathrm{nN}$ at scan line 1.  All other conditions are the same as for the trajectories plotted in Fig.~\ref{fig:sometrajectories}.
}
\label{fig:chaotic}
\end{figure}

In rare cases, particularly at very high load, where the nonlinearities are increased, periodic trajectories with a period longer than one lattice period as well as chaotic trajectories exist.
Examples of a period 6 periodic orbit and a chaotic orbit are displayed in Fig.~\ref{fig:chaotic}.
Nevertheless, even in these trajectories, the orientation remains roughly constant.

\begin{figure}
\includegraphics[width=8.4cm]{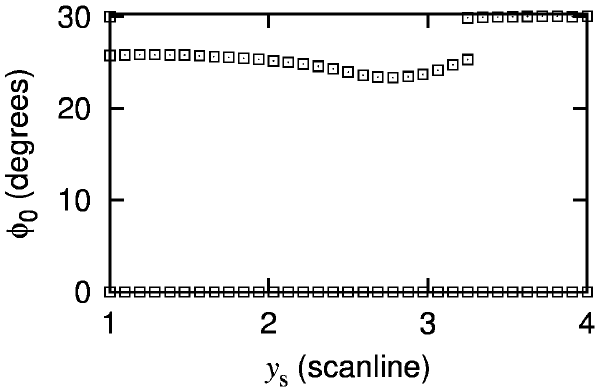}
\caption{
A bifurcation diagram of the stable periodic orbits as a function of the parameter $y$ for $N=24$ and $F_\mathrm{load}=20~\mathrm{nN}$.
The data was obtained by doing a large number of simulations with a wide range of initial conditions.
The plotted points are the set of final angles.
Clearly visible between scan lines 3 and 4 are the points at which the $\phi_0 \approx 26^\circ$ periodic orbit becomes unstable and the $\phi_0 \approx 30^\circ$ periodic orbit becomes stable.
\label{fig:scanline}
}
\end{figure}

\begin{figure}
\includegraphics[width=8.4cm]{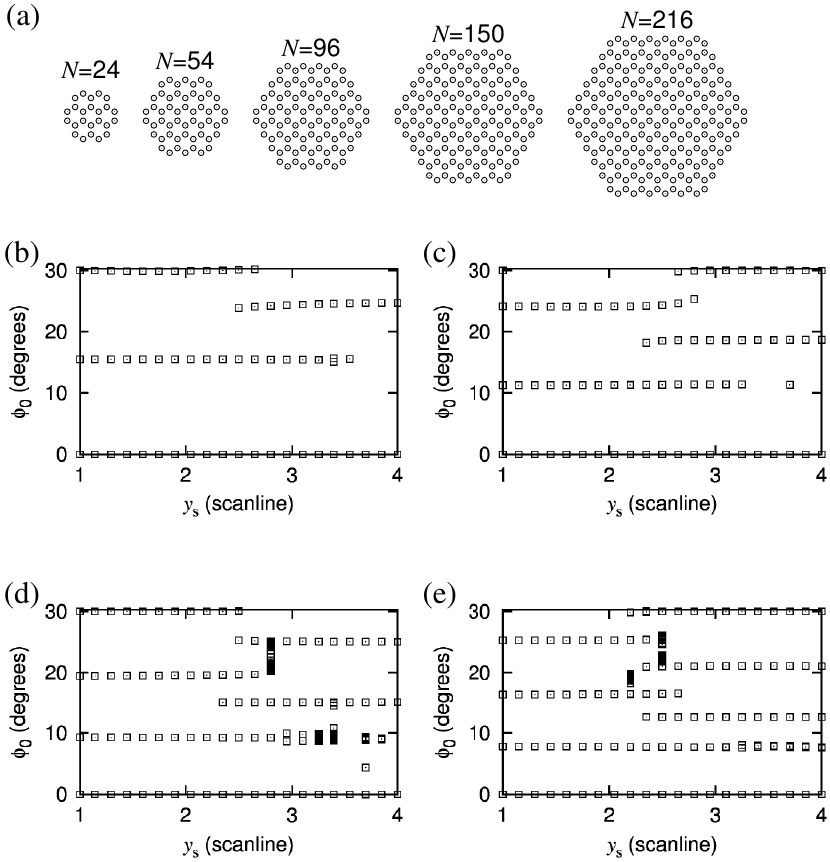}
\caption{
The plot of Fig.~\ref{fig:scanline} repeated for (a) various 6-fold symmetric flakes of (b) 54, (c) 96, (d) 150, and (e) 216 atoms.
Larger flakes have more stable periodic orbits.
The lone point in the bifurcation diagram for $N=96$ near scan line 4 at $\phi_0 \approx 12^\circ$ indicates that the $\phi_0 \approx 12^\circ$ periodic orbit is still stable there, but has such a small basin of attraction that the spacing between the initial conditions used to calculate this bifurcation diagram is not fine enough to detect it.
}
\label{fig:flakesizescanline}
\end{figure}

In Fig.~\ref{fig:scanline} the stable periodic orbits are plotted as a function of $y_\mathrm{s}$, ranging between scan line~1 and~4, for the system of Fig.~\ref{fig:sometrajectories}.
The commensurate periodic orbit at $\phi=0$ is always stable, regardless of the scan line.
Between scan lines~3 and~4 the incommensurate orbit at $\phi\approx 26^\circ$ becomes unstable, and the one at $\phi \approx 30^\circ$ becomes stable.

As the number of atoms increases, the interaction with the substrate becomes more complicated and the number of periodic orbits increases.
For square flakes on a square lattice, the number of periodic orbits increases linearly with the diameter of the flake \cite{tbp}.
In Fig.~\ref{fig:flakesizescanline}, bifurcation diagrams similar to Fig.~\ref{fig:scanline} are shown for flakes of different sizes.
The number of stable periodic orbits increases.
Additionally, there is a switch-over region around scan line 3, where the stable incommensurate orbits become unstable, and the unstable incommensurate orbits become stable.

Experimentally\cite{Dienwiebel2008} it was reported that the superlubric behavior of flakes of approximately 100~atoms lasted for about 40~scan lines or a distance $y_\mathrm{s}$ of about 7\AA, about 5~times the distance between scan lines~1 and~4.
This compares very well with the results for $N=96$, where we see that starting, for instance on scan line~4 and moving towards scan line~1, the flake rotates from the stable periodic orbit at $30^\circ$ to the one at $23^\circ$.
After this, due to the symmetry of the lattice, the scan moves back from scan line~1 to scan line~4, decreasing the mismatch angle to $19^\circ$ (which has lower energy than $30^\circ$).
After $3a\approx4.3$~\AA\ distance in the $y$ direction, the flake locks in the commensurate $\phi=0$ state.
In absence of thermal fluctuations the decay to the commensurate state is a geometric effect, depending only on the structure of the interaction.
Each periodic orbit leads to a different friction force, and so the observation of steps in the friction force going from one scan line to another, could be related to the size and symmetry of the flake.

\section{\label{sec:stability} Stability analysis and simplified models}

\begin{figure}
\includegraphics[width=8.4cm]{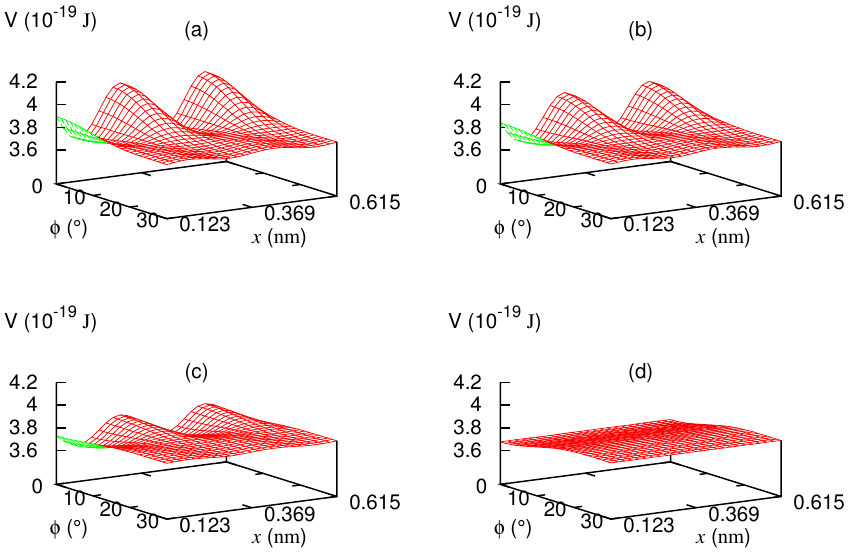}
\caption{
(Color online)
The potential energy $V(\vec{r},\phi)$ of a 24-atom flake as a function of the mismatch angle $\phi$ and position $x$ along the trajectory of the support, for constant $z$ at the average value belonging to a load of $20~\mathrm{nN}$, and $y$ corresponding to scan lines~(a)~1, (b)~2, (c)~3, and~(d)~4.
Due to the symmetries of the system given in Eqs.~(\ref{eq:symmsimple1}--\ref{eq:symmsimple3}), the dependence on $\phi$ is determined by the behavior between $0$ and $30^\circ$.
}
\label{fig:potentials}
\end{figure}

Although we consider the flake as a rigid object with only four degrees of freedom, the system is still too complicated to perform the stability analysis analytically.
However, a possible simplification is suggested by the shape of the potential energy $V(\vec{r}, \phi)$.
In Fig.~\ref{fig:potentials}, we show $V$ as a function of $x$ and $\phi$ for constant values of $z$ given by the average of the values found in simulations with a load of 20~nN, and $y$ defined by the four scan lines.
One can see that the potential is a periodic function of $x$ with an amplitude that depends on $\phi$.
Therefore, a good description of the system is provided by a simplified one-dimensional model with only two degrees of freedom: the position along the scan line, $x$, and the orientation, $\phi$.
This model is fully described by the viscous friction coefficient $\gamma$, support velocity $v_\mathrm{s}$, initial support position $x_{\mathrm{s}}^0$, mass $M$, moment of inertia $I$, and a simplified potential $V(x,\phi)$.
The essential dynamics of the system, the existence of commensurate and incommensurate sliding is preserved in the simplified model which we present in this section.

\subsection{equations of motion}

We write the equations of motion of the simplified system as a dynamical system of first-order differential equations,
\begin{align}
\label{eq:dotx} \dot{x} &= v_x~,\\
\label{eq:dotvx} M \dot{v}_x & = -\frac{\partial V(x,\phi)}{\partial x} - c (x-t v_{\mathrm{s}} - x_{\mathrm{s}~0})  - \gamma M v_x ~,\\
\label{eq:dotphi} \dot\phi& = \omega ~,\\
\label{eq:dotomega} I \dot\omega &= - \frac{\partial V({x},\phi)}{\partial \phi} -\gamma I \omega~.
\end{align}

Moreover, the symmetries of $V(\vec{r},\phi)$ in Eqs.~(\ref{eq:symmetries1}--\ref{eq:symmetries3}) imply that
\begin{align}
\label{eq:symmsimple1}
V(x,\phi) = V(x, \frac\pi3+\phi)~,\\
\label{eq:symmsimple2}
V(x,\phi) = V(x+l,\phi)~,\\
\label{eq:symmsimple3}
V(x,\phi) = V(-x,\pi-\phi)~,
\end{align}
where $l=a \sqrt{3}$.

\subsubsection{Specific potential}

A good representation of $V(x,\phi)$ for a given scan line ($y$ constant) is given by
\begin{align}
V(x,\phi) = U(\phi) + W(\phi) \cos\left(\frac{2 \pi x}{l}\right)~,
\label{eq:potentialUW}
\end{align}
where $U(\phi)$ and $W(\phi)$ are both smooth functions that represent the average value of the potential energy and the amplitude respectively.

\begin{figure}
\includegraphics[width=8.4cm]{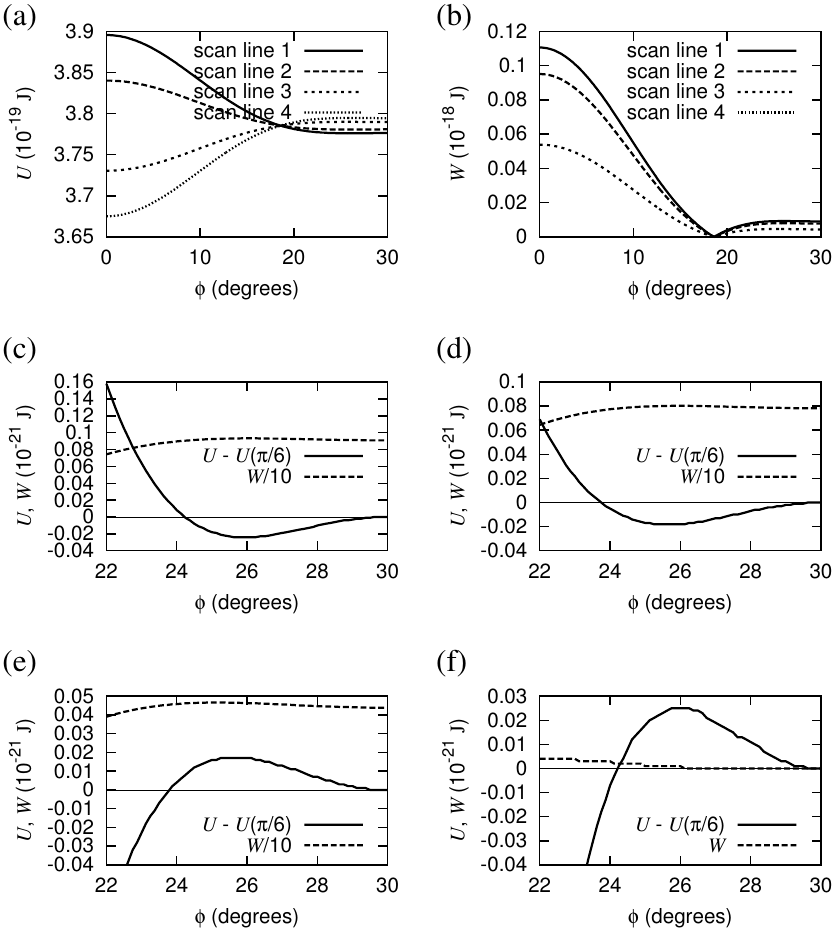}
\caption{
The (a) offset $U(\phi)$ and amplitude (b) $W(\phi)$ of the potential $V(x,\phi)$ as a function of $\phi$ for the same case displayed in Fig.~\ref{fig:potentials}.
The region near $\phi=30^\circ$ is enlarged separately for scan lines (c) 1, (d) 2, (e) 3, (f) 4.
$U$ and $W$ were obtained from a Fourier transform of $V$ with respect to $x$ over 492 points for each $\phi$.
The extrema of $U$ and $W$ coincide at $\phi=\phi_0$, which implies the existence of an invariant manifold $\phi=\phi_0, \omega=0$ with $\phi_0 = 0^\circ, 26^\circ$ or $30^\circ$.
}
\label{fig:UW}
\end{figure}

\begin{figure}
\includegraphics[width=8.4cm]{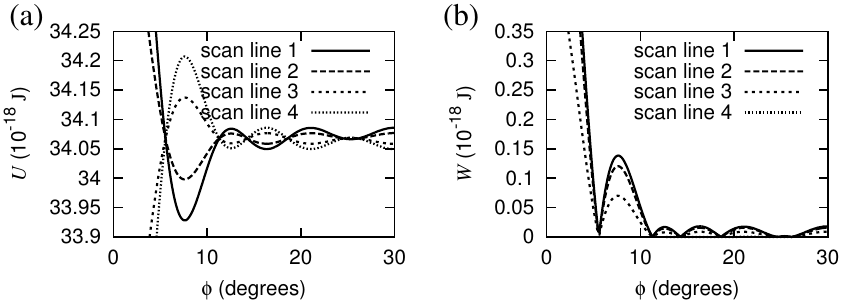}
\caption{
The (a) offset $U(\phi)$ and (b) amplitude $W(\phi)$ for a flake of 216 atoms.
The extrema of $U$ coincide with the maxima of $W$, and the nodes of $W$ correspond to a constant value of $U$.
There are more extrema than for the flake of $24$ atoms, and therefore more stable and unstable incommensurate periodic orbits.
}
\label{fig:UW216}
\end{figure}

The symmetries of the dynamics in Eqs.~(\ref{eq:symmsimple1}--\ref{eq:symmsimple3}) imply that
\begin{align}
U(\phi) = U(-\phi) = U\left(\frac\pi3+\phi\right)~,\\
W(\phi) = W(-\phi) = W\left(\frac\pi3+\phi\right) ~.
\end{align}
In turn, these equations imply that $U$ and $W$ have extrema in $\phi=\phi_0=0,\pi/6$.
In Figs.~\ref{fig:UW} and~\ref{fig:UW216}, $U$ and $W$, are shown for flakes of 24 and 216~atoms.
It is evident in Fig.~\ref{fig:UW} that there is an extremum of both $U$ and $W$ at $\phi=0$.
The structure of the other extremum, close to $30^\circ$ can be seen from the four enlargements.
Besides the extremum at $\phi=30^\circ$ for all scan lines there is another extremum of both $U$ and $W$ at about $26^\circ$.
In Fig.~\ref{fig:UW216}, for a larger flake, $U$ and $W$ have more extrema, but they still coincide.
In Ref.~\cite{tbp} it is shown for a simpler system, square flakes on a square lattice, that this is a general property: for square flakes on square lattices the extrema of $U$ and $W$ at any orientation coincide approximately for all flake sizes.

Since the torque, given by Eq.~(\ref{eq:dotomega}), vanishes for the values of $\phi$ that give extrema of $U$ and $W$ and $\omega=0$, these conditions define a two-dimensional invariant manifold of the dynamics.
The number of extrema of $U$ and $W$ and consequently the number of invariant manifolds grows with the size of the flake.

\subsection{stability\label{sec:substability}}

We consider a general potential $V(x,\phi)$ which has an invariant manifold at $\phi=\phi_0$, i.e.
\begin{align}
\label{eq:phi0}
\left. \frac{\partial V(x,\phi)}{\partial \phi}\right|_{\phi=\phi_0} = 0~,
\end{align}
for all $x$.
Now that we have identified the invariant manifold, we consider the dynamics in its vicinity in order to study the stability.

Near the invariant manifold, the torque is small, and so the time scales of $\phi$ and $\omega$, (Eqs.~(\ref{eq:dotphi}) and~(\ref{eq:dotomega})) are much longer than those of $x$ and $v_x$ (Eqs.~(\ref{eq:dotx}) and~(\ref{eq:dotvx})).
Because of this, for the purpose of investigating the stability of the dynamics near the invariant manifold, the torque can be replaced by its time average.
Note that this separation of time scales is only valid {\it near} the invariant manifold, namely if $\phi$ remains close to $\phi_0$ and $\omega$ is close to 0.

If the manifold is stable, then initial conditions close to it converge towards it.
We therefore consider the growth rates of small perturbations $\delta\phi$ and $\delta\omega$ of $\phi$ and $\omega$, the Lyapunov exponents.
From Eqs.~(\ref{eq:dotphi}) and~(\ref{eq:dotomega}) we find
\begin{align}
\dot{\delta\phi} & = \delta\omega~,\\
I \dot{\delta\omega} & = - \delta\phi\left.\frac{\partial}{\partial \phi} \left\langle \frac{\partial V({x},\phi) }{\partial \phi}\right\rangle_t \right|_{\phi=\phi_0} - \gamma I \delta\omega~.
\end{align}
The time average can be interchanged with the derivative with respect to $\phi$ because perturbations in $x$ and $\phi$ decouple to first order.
One may write
\begin{align}
\left(\begin{array}{l} \dot{\delta\phi}\\ \dot{\delta\omega} \end{array}\right)
&= 
\label{eq:matrix}
\left(\begin{array}{ll} 0&1\\ -\frac{1}{I} \left\langle\left.\frac{\partial^2  V({x},\phi)}{\partial \phi^2}\right|_{\phi=\phi_0}\right\rangle_t & -\gamma \end{array}\right)
\cdot
\left(\begin{array}{l} {\delta\phi}\\ {\delta\omega} \end{array}\right)\\
&= {\mathcal A} \cdot
\left(\begin{array}{l} {\delta\phi}\\ {\delta\omega} \end{array}\right) ~.
\end{align}
As the matrix ${\mathcal A}$ is constant, the Lyapunov exponents associated with perturbations in $\phi$ and $\omega$ are simply equal to its eigenvalues,
\begin{align}
\lambda_\pm = - \frac{1}{2} \gamma \pm \frac12 \sqrt{\gamma^2 - \frac{4}{I} \left\langle\left.\frac{\partial^2  V({x},\phi)}{\partial \phi^2}\right|_{\phi=\phi_0}\right\rangle_t }~.
\label{eq:lambda}
\end{align}
The invariant manifold is stable if all (in this case 2) Lyapunov exponents associated with perturbations of it have real components smaller than 0.

As the real component of the square root in Eq.~(\ref{eq:lambda}) is positive or 0, $\lambda_-\leq\lambda_+$ is the smallest Lyapunov exponent (i.e. has the smallest real component).
For stability analysis it therefore suffices to consider $\lambda_+$.
If the argument of the square root in Eq.~(\ref{eq:lambda}) is smaller than $\gamma^2$, then the real components of both $\lambda_-$ and $\lambda_+$ are negative.
This is the case if
\begin{align}
\label{eq:requirement}
\left\langle\left.\frac{\partial^2  V({x},\phi)}{\partial \phi^2}\right|_{\phi=\phi_0}\right\rangle_t >0~,
\end{align}
i.e., the time-average of the potential energy must be at a minumum.

Using Eq.~(\ref{eq:potentialUW}), Eq.~(\ref{eq:requirement}) can be rewritten to read
\begin{align}
\frac{\partial^2 U(\phi)}{\partial \phi^2} + \left.\frac{\partial^2W(\phi)}{\partial\phi^2}\right|_{\phi=\phi_0} \left\langle \cos\left(\frac{2 \pi x}{l}\right)\right\rangle_{t,\phi=\phi_0} > 0~.
\end{align}
The stability thus depends on $U$ and $W$, and how much time the particle spends near the minima of the potential, where the cosine is negative.

In stick-slip motion, the particle spends most of its time in the minima of the potential, i.e. where the cosine is smaller than 0 (see Fig.~\ref{fig:sometrajectories}).
If the motion is truly superlubric, then the particle spends about the same time in the minima as it does in the maxima.
If the motion is nearly superlubric, then the particle spends most of its time in the minima.
Hence, for realistic cases, $\langle \cos\rangle_t <0$.

If the offset of the potential, $U(\phi)$, has a minimum at $\phi_0$ it contributes positively towards the stability.
Similarly, if the amplitude $W(\phi)$ is at a maximum at $\phi_0$, because the first derivative is multiplied by a negative number, $\langle \cos\rangle_t$, it enhances the stability.
A minimum of $U$ and maximum of $W$ therefore always lead to stability, whereas a maximum of $U$ and minimum of $W$ always leads to instability.
If both are at a maximum, or both are at a minimum at $\phi_0$, then the stability is not directly obvious.

\subsubsection{comparison with simulations\label{sec:stabilityscanline}}

The analysis of Sec.~\ref{sec:substability} compares very well with the results of numerical simulations at $T=0~\mathrm{K}$.
The stability of the commensurate and incommensurate states can be determined by looking at the behavior of the average potential energy $U$ and amplitude $W$, shown in Figs.~\ref{fig:UW} and~\ref{fig:UW216}.

We examine first the 24-atom system of Figs.~\ref{fig:sometrajectories} and~\ref{fig:scanline}, for which $U$ and $W$ are reported in Fig.~\ref{fig:UW}.
For scan line~1 and~2, the minimum of $U$ at $\phi=26^\circ$ coincides with a maximum of $W$, and is therefore stable.
This is consistent with the simulation results for scan lines~1 and~2, shown in Fig.~\ref{fig:scanline}, where we see a stable orbit at $26^\circ$.
At these scan lines, for $\phi=30^\circ$ $U$ has a maximum and $W$ has a minimum, leading to instability.
At $\phi=0$, there is a maximum in $U$, but also in $W$.
However, the second derivatives of $U$ and $W$ are very nearly the same apart from the sign, and  $\langle \cos \rangle_t $ increases with decreasing amplitude, so $\langle \partial^2 V/\partial \phi^2 \rangle_t$ is positive, and the incommensurate state is stable.

For scan lines~3 and~4, at $0^\circ$ the minimum of $U$ coincides with a maximum in $W$, leading to a stable commensurate state.
Similarly, at scan line~4, the incommensurate state at $\phi=30^\circ$ is stable, while the state at $\phi=26^\circ$ is unstable.
For scan line~3, the stability of the incommensurate states is more complicated, as $U$ and $W$ both have maxima around $\phi=26^\circ$ and minima at $\phi=30^\circ$.
However, the second derivatives of $U$ and $W$ for both states are approximately the same, with opposite sign.
Additionally, the amplitude for both states is approximately the same, so $\langle \cos \rangle_t$ should be the same as well.
The crucial quantity for stability, $\langle \partial^2 V/\partial \phi^2 \rangle_t$, should therefore be nearly the same for the two states, except for the sign, which is opposite.
One of the incommensurate states is therefore stable, while the other is unstable, though from $U$ and $W$ it is not directly clear which is which.
In Fig.~\ref{fig:sometrajectories}, the stable incommensurate state for scan line~3 is shown at $\phi=26^\circ$ and for scan line~4 at $\phi=30^\circ$.
The existence of a switch-over scan line can be seen in the simulation results in Fig.~\ref{fig:scanline}, and is clearly critical for all sizes, as shown in Fig.~\ref{fig:flakesizescanline}.
Its existence for any flake size can be demonstrated analytically for square flakes on square lattices \cite{tbp}.

In Fig.~\ref{fig:UW216}, $U$ and $W$ are plotted for a larger flake of 216 atoms.
Because of the larger size of the flake, $U$ and $W$ have more extrema and therefore there are more periodic orbits.
The stable periodic orbits in the simulations, shown in Fig.~\ref{fig:flakesizescanline} (bottom right), coincide with the extrema.
Their stability is also consistent with calculations based on $U$ and $W$.

\begin{figure*}
\includegraphics[width=17.4cm]{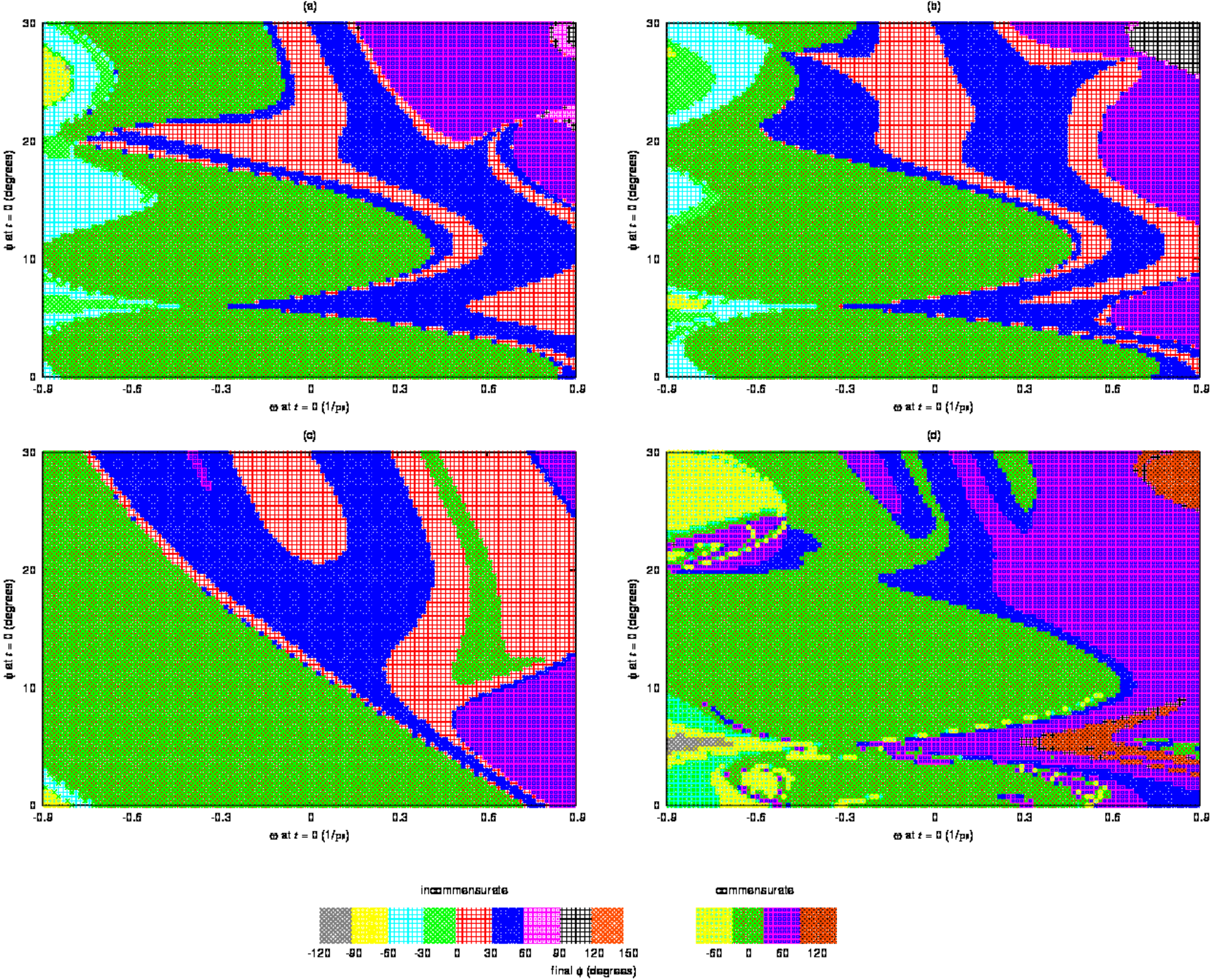}
\caption{
(Color online)
Cross sections of the phase space, including the basin of attraction of the incommensurate stable periodic orbits, which are at $\phi \approx 26^\circ,\omega=0$, for scan lines (a) 1, (b) 2, (c) 3, and $\phi\approx 30^\circ,\omega=0$ for (d) scan line 4.
The flake has 24 atoms and $F_\mathrm{load} = 20~\mathrm{nN}, v_\mathrm{s} = 32~\mathrm{m/s}$.
The final state of the flake is plotted as a function of the initial orientation and angular momentum.
The initial position and velocity have been chosen such that the stable incommensurate periodic orbit intersects with the cross section, in $\omega=0$.
The colours indicate to which periodic orbit the initial conditions converge:
red incommensurate $\phi_0\in \langle0^\circ, 30^\circ]$,
blue incommensurate $\phi_0\in [30^\circ, 60^\circ\rangle$,
purple with blue commensurate $\phi_0=60^\circ$,
purple incommensurate $\phi_0\in \langle 60^\circ, 90^\circ]$,
black incommensurate $\phi_0\in [90^\circ, 120^\circ\rangle$,
red with black commensurate $\phi_0=120^\circ$,
green with red commensurate $\phi_0=0^\circ$,
green incommensurate $\phi_0\in [-30^\circ, 0^\circ\rangle$,
cyan incommensurate $\phi_0\in \langle -60^\circ, -30^\circ]$,
yellow with cyan commensurate $\phi_0=-60^\circ$.
yellow incommensurate $\phi_0\in [-90^\circ, 60^\circ\rangle$.
grey incommensurate $\phi_0\in \langle -120^\circ, -90^\circ]$.
The incommensurate periodic orbit at $\phi_0\approx 30^\circ$ is indicated in blue, as it visits both $\langle0^\circ, 30^\circ]$ and $[30^\circ, 60^\circ\rangle$ in one period.
}
\label{fig:crossection}
\end{figure*}

\section{Robustness of the superlubric sliding\label{sec:robustness}}

The analysis presented in Sec.\ref{sec:stability} shows that incommensurate (superlubric) sliding may exist.
However, the existence of stable incommensurate periodic orbits does not necessarily mean that they can be easily observed in experiments.
The conditions which lead to stability may not be experimentally accessible.
Furthermore, the stability may be very weak, causing very slow convergence towards the periodic orbit, or the basins of attraction of the incommensurate periodic orbits (the set of initial conditions that converge towards them) may be small.
In this section we examine separately the robustness of the incommensurate superlubric solutions against several types of perturbations.

\subsection{temperature}

\begin{figure}
\includegraphics[width=8.4cm]{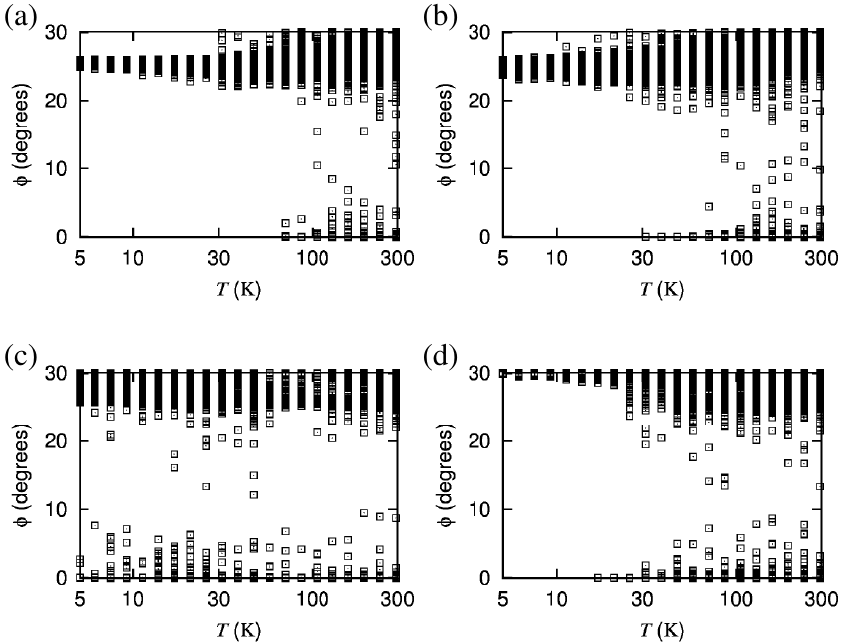}
\caption{
The final orientation of a 24-atom flake (mapped onto the interval $[0^\circ,30^\circ]$) which was initially in the stable incommensurate periodic orbit is plotted as a function of temperature after a long, but finite time, with $F_\mathrm{load} = 20~\mathrm{nN}, v_\mathrm{s} = 32~\mathrm{m/s}$ for scan lines (a)~1, (b)~2, (c)~3, and (d)~4.
For every temperature, 250 realisations are plotted.
Enough time has elapsed for the system to decay to the static state distribution.
}
\label{fig:temperature}
\end{figure}

In Fig.~\ref{fig:crossection}, we show slices of the phase space which contain the stable incommensurate periodic orbits of the full three-dimensional system.
For each scan line, we investigate the basin of attraction by performing numerical simulations at $T=0$ and looking at the asymptotic state of the flake as a function of the initial orientation and angular momentum.
If the basin of attraction is small, the periodic orbits can easily be destroyed by thermal fluctuations, which bring the system outside the basin of attraction of the incommensurate orbits, and into that of the commensurate orbit.
The range of initial angular velocities plotted in Fig.~\ref{fig:crossection} is $3\sqrt{k_\mathrm{b} T_\mathrm{r}/M}$, where $T_\mathrm{r}$ is room temperature, 293~K, and $k_\mathrm{b}$ is Boltzmann's constant.
This is roughly the range that is thermally accessible at room temperature.
The basin of attraction of the incommensurate periodic orbits is smaller than this range, indicating that at room temperature thermal fluctuations may perturb the incommensurate state sufficiently to cause it to decay to the commensurate state, which has lower energy.
Especially scan line~3, with its weak stability and scan line~4, at which the incommensurate state only has a small basin of attraction (as is shown in Fig.~\ref{fig:crossection}), are very sensitive to thermal fluctuations.

To examine the effect of temperature explictly we conduct Langevin simulations with temperatures ranging from 5 to 300~$\mathrm{K}$.
Starting from initial conditions on the incommensurate periodic orbit, simulated systems were subjected to thermal fluctuations for a period of about 100 lattice periods and the final angle was recorded.
The results are plotted in Fig.~\ref{fig:temperature}.
At scan line~3 the incommensurate state is the least robust against temperature and the incommensurate state decays already at 5~K.
However, thermal fluctuations are not the only source of energy in this system, because the moving support drives the flake at velocities that are not negligeable compared to thermal velocities, therefore supplying amounts of energy significant compared to $k_\mathrm{B} T$.
The effect of temperature is thus possibly overestimated in these simulations.

\subsection{scan line}

From the size of the basins of attraction in Fig.~\ref{fig:crossection} and the robustness against thermal fluctuations, displayed in Fig.~\ref{fig:temperature}, it can be seen that the robustness of the incommensurate states in this system depends strongly on the scan line.
For the system in the figures, the incommensurate periodic orbit is the least robust for scan lines~3 and 4.
From Fig.~\ref{fig:UW} it can be seen that the minimum of $U$ near $\phi=30^\circ$ is shallow and the amplitude $W$, especially in the case of scan line~4, is small.
The latter is a consequence of the symmetries of the hexagonal lattice.

As discussed in Sec.~\ref{sec:per_orbits}, the different stability and instability of the incommensurate states at different scan lines can lead to the disappearance of superlubricity after an initial superlubric period in experiments which explore more than one scan line.
Additionally, the weak stability of the incommensurate states, and associated low robustness against thermal fluctuations, near scan lines~3 and~4 makes superlubric states less likely to persist in such experiments.

\subsection{flake size}

\begin{figure}
\includegraphics[width=8.4cm]{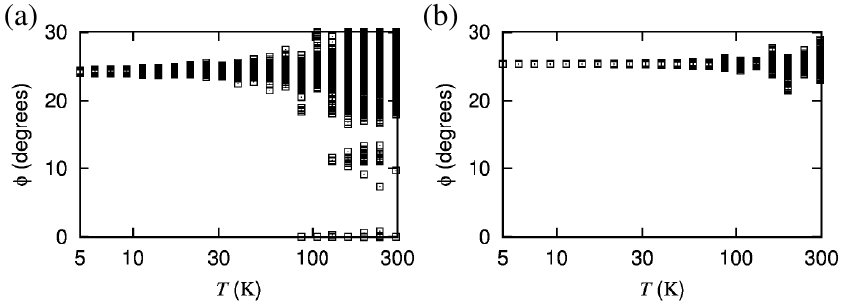}
\caption{
The plot of Fig.~\ref{fig:temperature} for scan line~2 repeated for flakes of (a) 96 and (b) 216 atoms.
The stable periodic orbits of large flakes are more robust against temperature, because the moment of inertia grows as $N^2$.
\label{fig:flakesizetemp}
}
\end{figure}

As the number of atoms in the flake increases
the moment of inertia increases with $N^2$.
This means that the orientation and angular velocity of larger flakes are less sensitive to thermal fluctuations and other disruptions.
By comparing Fig.~\ref{fig:flakesizetemp} to Fig.~\ref{fig:temperature}
we see that the incommensurate periodic orbit of the flake with 216 atoms is more robust against thermal fluctuations, and survives even at room temperature.

It is interesting to note that Bonelli et al.\cite{Bonelli2009}, who consider flexible graphite flakes, found that larger flakes interact more weakly with the substrate than one would expect from rigid flakes.
At the edges, the flake bends towards the substrate, and thus the atoms at the edge of the flake dominate the interaction.
However, in Ref.~\cite{Bonelli2009}, no analysis of the stability of superlubricity was possible, as the coupling between the cantilever and flake was chosen in such a way as to impose a preferred orientation.

\subsection{support velocity}

\begin{figure}
\includegraphics[width=8.4cm]{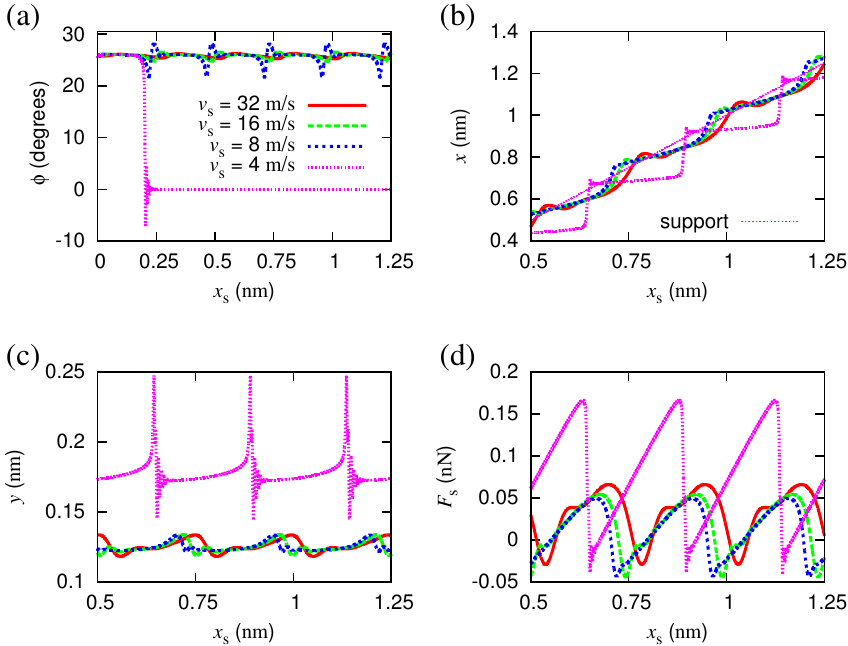}
\caption{
(Color online) The (a) orientation $\phi$, positions (b) $x$ and (c) $y$, and (d) friction $F_\mathrm{s}$ as a function of support position $x_\mathrm{s}$ for various support velocities and $N=24$, $F_\mathrm{load}=20~\mathrm{nN}$, scan line~2.
As the velocity decreases, the fluctuations in $\phi$ and $y$ increase and the system behaves less one-dimensionally.
For sufficiently low $v_\mathrm{s}$, the system can no longer be described by the simplified model.
}
\label{fig:supportvelocity}
\end{figure}

At high support velocity, the motion of the flake is less sensitive to the detailed structure of the substrate.
The dynamics in the $y$ and $z$ direction are relatively fast compared to the dynamics of the rotation, and therefore their effects on the orientation of the flake average out.
At lower support velocities, motion in the $y$ and $z$ direction becomes more relevant and can reduce the size of the basin of attraction of the incommensurate periodic orbits, or even destroy the stability completely.
In Fig.~\ref{fig:supportvelocity} trajectories are plotted for the same flake at different support velocities.
As the velocity decreases, the flake becomes more sensitive to fluctuations and therefore $\phi$ (top left) and $y$ (bottom left) fluctuate more.
At sufficiently low velocities, the incommensurate periodic orbit is no longer stable, and the flake rotates to the commensurate orientation with stick-slip motion (top right) and high friction (bottom right).
A stronger coupling between the flake and cantilever would reduce the fluctuations in the $y$ direction, and allow the stability of the incommensurate state to persist to lower support velocities.

\subsection{load\label{sec:load}}

\begin{figure}
\includegraphics[width=8.4cm]{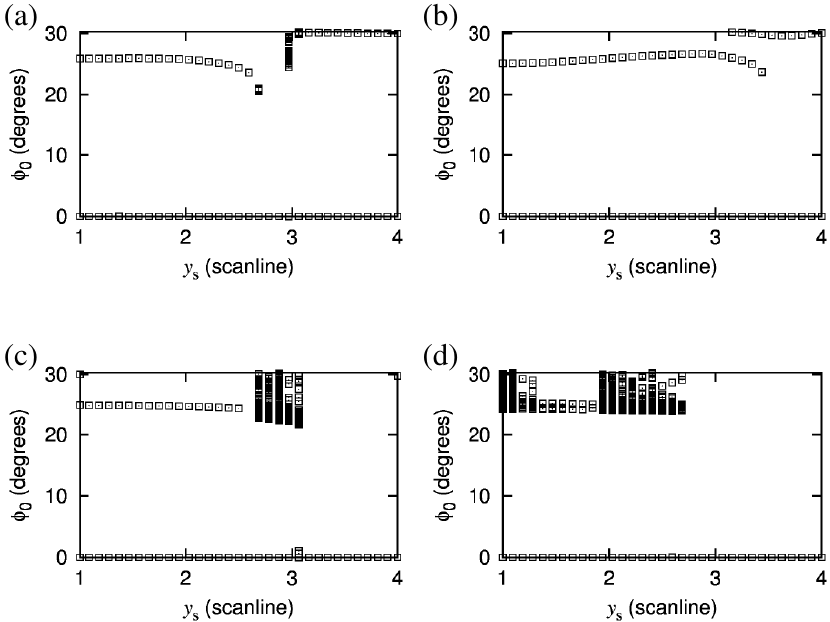}
\caption{
The plot of Fig.~\ref{fig:scanline} repeated for load force equal to (a) $0~\mathrm{nN}$, (b) $10~\mathrm{nN}$, (c) $30~\mathrm{nN}$, and (d) $40~\mathrm{nN}$.
}
\label{fig:load}
\end{figure}

The load force exerted by the cantilever on the flake pushes it into the substrate.
This affects not only the corrugation, but also the shape of the potential to which the flake is subjected.
Consequently, for different load forces, the behavior of $U$ and $W$ is different, and so the stability of incommensurate periodic orbits may change.
In Fig.~\ref{fig:load} bifurcation diagrams similar to the one in Fig.~\ref{fig:scanline} are shown for different load forces.
When the load is very high, the interaction between the flake and substrate is changed qualitatively, and for the region near scan lines~3 and~4 the incommensurate periodic orbit disappears.
The simulations of Ref.~\cite{Bonelli2009} were performed using load forces of about 100~nN, and indeed, no superlubric behavior was observed.
Bonelli et al. also performed a few simulations at lower loads for 24 atom flakes, but imposed mismatch angles near 0$^\circ$ and 15$^\circ$ on the flake, thus eradicating the incommensurate periodic orbits near 30$^\circ$.

At high load force periodic trajectories with periods longer than one lattice period and chaotic trajectories exist.
Two such trajectories are shown in Fig.~\ref{fig:chaotic}.
These trajectories still have roughly constant orientation, because the invariant manifold is still stable.
It is the motion on the invariant manifold itself that has a longer period or is chaotic.

\subsection{choice of potential}

\begin{figure}
\includegraphics[width=8.4cm]{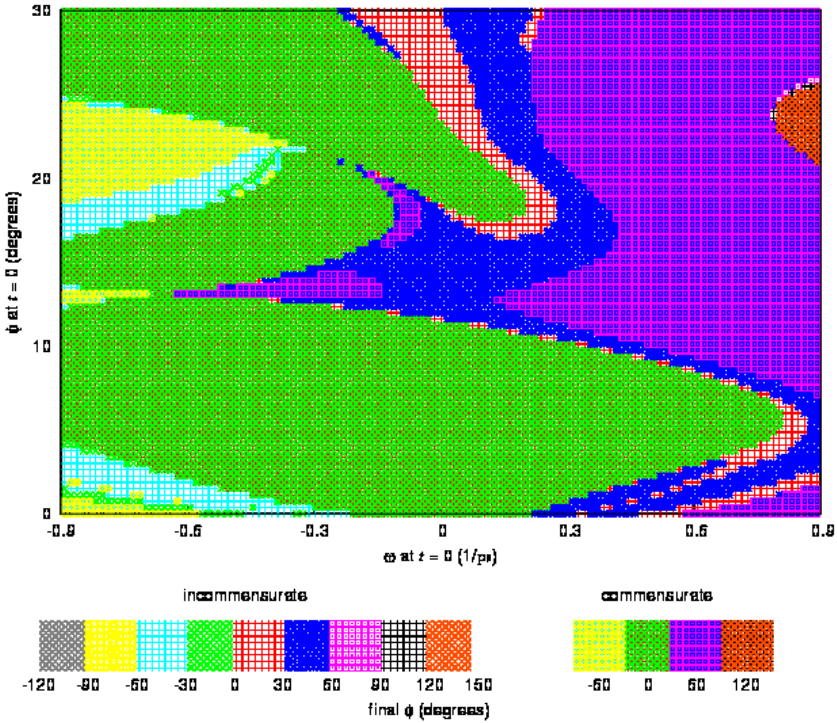}
\caption{(Color online)
The cross section for scan line 3 in Fig.~\ref{fig:crossection} repeated with the commonly used 2D interaction potential.
The basin of attraction is different in shape and size.
}
\label{fig:3dvs2d}
\end{figure}

\begin{figure}
\includegraphics[width=8.4cm]{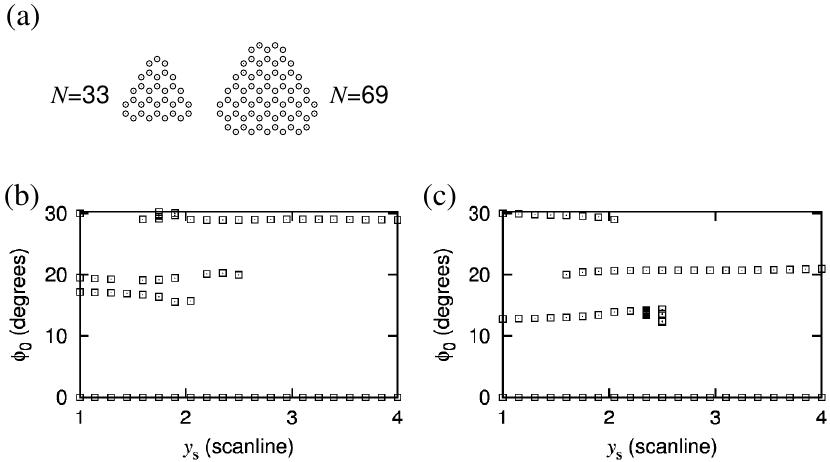}
\caption{
The plot of Fig.~\ref{fig:scanline} repeated for (a) various 3-fold symmetric flakes of (b) 33 and (c) 69 atoms.
}
\label{fig:flakeshape}
\end{figure}

Very often, for friction, the potential corrugation is represented as a two-dimensional profile in the $xy$ plane.
In this representation, the load can only be included by scaling the potential.
Figs.~\ref{fig:UW} and~\ref{fig:UW216} would therefore look the same but only scaled, regardless of load, which implies that the stable incommensurate orbits would remain stable for any load.
This is not the case for the 3-d potential used here, as can be seen from Fig.~\ref{fig:load}.

However, in a fully three-dimensional problem, the effect of load is not simply a rescaling of the amplitude.
We compare our results obtained with a three-dimensional potential to the one obtained with the two-dimensional potential of reference\cite{Dienwiebel2008}.
We find (Fig.~\ref{fig:3dvs2d}) that the cross section has qualitatively the same features, but a significantly different size of the basin of attraction.

\subsection{symmetry of the flakes}

In experimental conditions, it cannot be guaranteed that the flakes are exactly hexagonal.
In Fig.~\ref{fig:flakeshape} the bifurcation diagrams of Figs.~\ref{fig:scanline} and~\ref{fig:flakesizescanline} have been repeated for three-fold symmetric flakes of two different sizes.
The results are similar to those of the hexagonal flakes, though somewhat distorted.

\section{Conclusions\label{sec:conclusions}}

In this paper, we have examined the possibility of realising conditions for superlubric sliding without rotation and locking of graphite flakes on graphite.
By means of a simplified analytical model, validated by our numerical simulations, we have shown that incommensurate periodic orbits with low friction can be stable.
Furthermore, we have investigated the robustness of the superlubric sliding against changes in several conditions and quantities: temperature, scan line, flake size, support velocity, load, and asymmetry.

Our results show that some scan lines, where the center of mass moves along a row of atoms of the substrate, are detrimental to the stability of superlubric sliding and lead to rotation of the flake as found in Ref.~\cite{Dienwiebel2008}.
Conversely, superlubric sliding is favored by larger flakes, higher velocities than in AFM, and low temperature.
Our calculations suggest that in an experiment where different scan lines are explored successively the locking would occur gradually via intermediate periodic orbits.
For a flake of about 100 atoms, this should occur in 4 steps.
As the friction force for each periodic orbit is different, this could perhaps be used as a method for characterising the flake.

\begin{acknowledgments}

ASW's work is financially supported by a Veni grant of Netherlands Organisation for Scientific Research (NWO).
AF's work is part of the research programme of the Foundation for Fundamental Research on Matter (FOM), which is financially supported by the Netherlands Organisation for Scientific Research (NWO).
AF would like to thank A.E.~Filippov, M.~Urbakh, and J.W.M.~Frenken for discussions.

\end{acknowledgments}


\begin{thebibliography}{99}
\bibitem{Dienwiebel2004} M.~Dienwiebel, G.S.~Verhoeven, N.~Pradeep, J.W.M.~Frenken, J.A.~Heimberg, and H.W.~Zandbergen, Phys.~Rev.~Lett. {\bf 92}, 126101 (2004).
\bibitem{Dienwiebel2008}A.E.~Filippov, M.~Dienwiebel, J.W.M.~Frenken, J.~Klafter, M.~Urbakh, Phys.\ Rev.\ Lett. {\bf 100},  046102 (2008).
\bibitem{FK} Ya.I.~Frenkel and T.A.~Kontorova, Zh.~Eksp.~Teor.~Fiz. {\bf 8}, 89 (1938).
\bibitem{Aubry} M.~Peyrard and S.~Aubry, J.~Phys.~C {\bf 16}, 1593 (1983); S. Aubry and L. de Seze, Festk\"orperprobleme XXV, 59 (1985).
\bibitem{Shinjo} K.~Shinjo, M.~Hirano, Surf Sci. {\bf 283}, 473 (1993).
\bibitem{Hirano}M.~Hirano, K.~Shinjo, R.~Kaneko and Y.~Murata, Phys.~Rev.~Lett. {\bf 78}, 1448 (1997).
\bibitem{Consoli}  L.~Consoli, H.J.F.~Knops, A.~Fasolino, Phys.~Rev.~Lett.~{\bf 85}, 302 (2000); Phys.~Rev.~E{\bf 64}, 016601 (2001).
\bibitem{Socoliuc} A. Socoliuc, R. Bennewitz, E. Gnecco, E. Meyer, Phys.~Rev.~Lett. {\bf 92}, 134301 (2004).
\bibitem{Weiss} M.~Weiss, F.J.~Elmer, Phys.~Rev.~{\bf 53}, 7539 (1996).
\bibitem{Los2003} J.H.~Los, A.~Fasolino, Phys.~Rev.~B {\bf 68}, 024107 (2003).
\bibitem{vverletspullie} See e.g. C.~Fusco, PhD thesis, Radboud Universiteit Nijmegen (2005) \url{http://repository.ubn.ru.nl/bitstream/2066/32530/1/32530_fricandid.pdf}.
\bibitem{tbp} To be published.
\bibitem{Bonelli2009} F.~Bonelli, N.~Manini, E.~Cadelano, L.~Colombo, Eur.~Phys.~J.~B {\bf 70}, 449-459 (2009).
\end{thebibliography}
\end{document}